\documentclass[
	a4paper, % Paper size, use either a4paper or letterpaper
	10pt, % Default font size, can also use 11pt or 12pt, although this is not recommended
	unnumberedsections, % Comment to enable section numbering
	twoside, % Two side traditional mode where headers and footers change between odd and even pages, comment this option to make them fixed
]{LTJournalArticle}

\usepackage{glossaries}

\newacronym{udma}{\textmu DMA}{micro direct memory access}
\newacronym{abi}{ABI}{application binary interface}
\newacronym{ace}{ACE}{AXI Coherent Extensions}
\newacronym{alu}{ALU}{arithmetic logic unit}
\newacronym{amba}{AMBA}{Advanced Microcontroller Bus Architecture}
\newacronym{amo}{AMO}{atomic memory operation}
\newacronym{aot}{AOT}{ahead-of-time}
\newacronym{apb}{APB}{Advanced Peripheral Bus}
\newacronym{api}{API}{application programming interface}
\newacronym{asic}{ASIC}{application-specific integrated circuit}
\newacronym{axi}{AXI}{Advanced eXtensible Interface}
\newacronym{b5g}{B5G}{Beyond-5G}
\newacronym{blas}{BLAS}{Basic Linear Algebra Subprograms}
\newacronym{cas}{CAS}{compare-and-swap}
\newacronym[longplural={core complexes}]{cc}{CC}{core complex}
\newacronym{cgra}{CGRA}{coarse-grained reconfigurable architecture}
\newacronym{cmos}{CMOS}{complementary metal-oxide-semiconductor}
\newacronym{cnn}{CNN}{convolutional neural network}
\newacronym{cpu}{CPU}{central processing unit}
\newacronym{csr}{CSR}{control and status register}
\newacronym{dbt}{DBT}{dynamic binary translation}
\newacronym{dct}{DCT}{discrete cosine transform}
\newacronym{dit_adj}{DIT}{decimation-in-time}
\newacronym{dif_adj}{DIF}{decimation-in-frequency}
\newacronym{dlp}{DLP}{data-level parallelism}
\newacronym{dma}{DMA}{direct memory access}
\newacronym{dram}{DRAM}{dynamic random-access memory}
\newacronym{dse}{DSE}{design space exploration}
\newacronym{dsl}{DSL}{domain-specific language}
\newacronym{dsp}{DSP}{digital signal processing}
\newacronym{elf}{ELF}{Executable and Linkable Format}
\newacronym{fdsoi}{FD-SOI}{fully depleted silicon-on-insulator}
\newacronym{fft}{FFT}{Fast Fourier transform}
\newacronym{dft}{DFT}{Discrete Fourier transform}
\newacronym{fifo}{FIFO}{first in, first out}
\newacronym{fpga}{FPGA}{field-programmable gate array}
\newacronym{fpu}{FPU}{floating-point unit}
\newacronym{fsm}{FSM}{finite-state machine}
\newacronym{gemm}{GEMM}{general matrix multiplication}
\newacronym{gpgpu}{GPGPU}{general-purpose computing on a \gls{gpu}}
\newacronym{gpu}{GPU}{graphics processing unit}
\newacronym{hart}{hart}{hardware thread}
\newacronym{hbm}{HBM}{High Bandwidth Memory}
\newacronym{hdl}{HDL}{hardware description language}
\newacronym{hero}{HERO}{Heterogeneous Embedded Research Platform}
\newacronym{hpc}{HPC}{high-performance computing}
\newacronym{ilp}{ILP}{instruction level parallelism}
\newacronym{ipc}{IPC}{instructions per cycle}
\newacronym{iot}{IoT}{Internet of Things}
\newacronym{ipu}{IPU}{integer processing unit}
\newacronym{ir}{IR}{intermediate representation}
\newacronym{isa}{ISA}{instruction set architecture}
\newacronym{issr}{ISSR}{indirection stream semantic register}
\newacronym{ita}{ITA}{2Integer Transformer Accelerator}
\newacronym{jit}{JIT}{just-in-time}
\newacronym{llc}{LLC}{last-level cache}
\newacronym{llm}{LLM}{large language model}
\newacronym{lrsc}{LRSC}{load-reserved/store-conditional}
\newacronym{lr}{LR}{load-reserved}
\newacronym{lsu}{LSU}{load-store unit}
\newacronym{mac}{MAC}{multiply–accumulate}
\newacronym{mimd}{MIMD}{multiple instruction, multiple data}
\newacronym{mmu}{MMU}{memory management unit}
\newacronym[longplural={networks-on-chip}]{noc}{NoC}{network-on-chip}
\newacronym{numa}{NUMA}{non-uniform memory access}
\newacronym{ppa}{PPA}{power, performance, and area}
\newacronym{pc}{PC}{program counter}
\newacronym{pe}{PE}{processing element}
\newacronym{pl}{PL}{programmable logic}
\newacronym{pmca}{PMCA}{programmable manycore accelerator}
\newacronym{pulp}{PULP}{Parallel Ultra Low Power}
\newacronym{qlr}{QLR}{queue-linked register}
\newacronym{raw}{RAW}{read-after-write}
\newacronym{rom}{ROM}{read-only memory}
\newacronym{rmw}{RMW}{read–modify–write}
\newacronym{rob}{ROB}{reorder buffer}
\newacronym{ro}{RO}{read-only}
\newacronym{rtl}{RTL}{register-transfer level}
\newacronym{rvv}{RVV}{RISC-V Vector Extension}
\newacronym{sbt}{SBT}{static binary translation}
\newacronym{scm}{SCM}{standard cell memory}
\newacronym{sc}{SC}{store-conditional}
\newacronym{sdf}{SDF}{Standard Delay Format}
\newacronym{simd}{SIMD}{single instruction, multiple data}
\newacronym{simt}{SIMT}{single instruction, multiple thread}
\newacronym{sm}{SM}{streaming multiprocessor}
\newacronym{soc}{SoC}{system-on-chip}
\newacronym[longplural={scratchpad memories}]{spm}{SPM}{scratchpad memory}
\newacronym{sram}{SRAM}{static random-access memory}
\newacronym{ssa}{SSA}{static single assignment}
\newacronym{ssr}{SSR}{stream semantic register}
\newacronym{tcdm}{TCDM}{tightly-coupled data memory}
\newacronym{tlp}{TLP}{thread-level parallelism}
\newacronym{tpu}{TPU}{Tensor Processing Unit}
\newacronym{vpu}{VPU}{vector processing unit}
\newacronym{vliw}{VLIW}{very long instruction word}
\newacronym{vnb}{VNB}{von Neumann bottleneck}
\newacronym{war}{WAR}{write-after-read}
\newacronym{waw}{WAW}{write-after-write}
\newacronym{xr}{XR}{extended reality}

\runninghead{} % A shortened article title to appear in the running head, leave this command empty for no running head

\footertext{\textit{RISC-V Summit Europe, Paris, 12-15th May 2025}} % Text to appear in the footer, leave this command empty for no footer text

\title{MemPool Flavors: Between Versatility and Specialization in a RISC-V Manycore Cluster}

\author{%
	Sergio Mazzola\textsuperscript{1}\thanks{Corresponding author: \href{mailto:smazzola@iis.ee.ethz.ch}{\tt smazzola@iis.ee.ethz.ch}}, Yichao Zhang\textsuperscript{1}, Marco Bertuletti\textsuperscript{1}, Diyou Shen\textsuperscript{1}, and Luca Benini\textsuperscript{1,2}
}

\date{\footnotesize\textsuperscript{\textbf{1}}Integrated Systems
Laboratory (IIS), Swiss Federal Institute of Technology (ETH Zürich), Switzerland\\ \textsuperscript{\textbf{2}}Department of Electrical, Electronic and Information Engineering (DEI), University of Bologna, Italy}
\renewcommand{\maketitlehookd}{%
	\begin{abstract}
        \noindent As computational paradigms evolve, applications such as attention-based models, wireless telecommunications, and computer vision impose increasingly challenging requirements on computer architectures: significant memory footprints and computing resources are demanded while maintaining flexibility and programmability at a low power budget.
        Thanks to their advantageous trade-offs, shared-L1-memory clusters have become a common building block of massively parallel computing architectures tackling these issues.
        MemPool is an open-source, RISC-V-based manycore cluster scaling up to 1024 \glspl{pe}. MemPool offers a scalable, extensible, and programmable solution to the challenges of shared-L1 clusters, establishing itself as an open-source research platform for architectural variants covering a wide trade-off space between versatility and performance. As a demonstration, this paper compares the three main MemPool flavors, Baseline MemPool, Systolic MemPool, and Vectorial MemPool, detailing their architecture, targets, and achieved trade-offs.
        \vspace{-2mm}
	\end{abstract}
}

\begin{document}

\maketitle % Output the title section

%----------------------------------------------------------------------------------------
%	ARTICLE CONTENTS
%----------------------------------------------------------------------------------------

%%%%%%%%%%%%%%%%%%%%%%%%%%%
% INTRODUCTION
%%%%%%%%%%%%%%%%%%%%%%%%%%%

\section{Introduction}
\vspace{-2mm}
Manycore systems are a class of parallel computer architectures ranging from tens to thousands of \glspl{pe}. They cover a wide trade-off space among versatility, performance, and energy efficiency, where their position is determined by the specialization of the \glspl{pe},
%(from general-purpose, independent cores to domain-specific accelerators)
their interconnect topology,
%(from all-to-all to neighboring-based)
and the nature of the memory subsystem.
%(from flat \glspl{spm} with \gls{numa} to deep cache hierarchies).
%
Their massive parallelism makes manycore systems particularly suitable for many transformative technologies such as telecommunications, with \gls{b5g} and 6G, artificial intelligence, with \glspl{llm}, and computer vision, with \gls{xr}.
Along with the high computational demands of these target applications, however, massive parallelism also introduces challenges such as complex programming models and communication bottlenecks.

MemPool is an open-source, RISC-V-based manycore cluster scaling up to 1024 \glspl{pe} sharing a pool of multi-banked L1 \gls{spm} through a hierarchical, low-latency interconnect~[1]. MemPool's \glspl{pe} are small, memory-latency-tolerant 32-bit RISC-V cores supporting custom \gls{isa} extensions~[2].
MemPool strikes a new trade-off between general-purpose, inefficient systems and fully specialized, high-performance architectures.
The simple, full-fledged cores allow programmability and versatility, with their flexible datapath enhancing performance. Moreover, the hierarchical interconnect provides the cores with low-latency and energy-efficient access to any address of the shared memory, decoupling the system from domain-specific dataflows.

Since its first developments in 2020, MemPool’s flexibility has proven to extend beyond just architectural versatility. Not only is it an efficient, general-purpose system, but it has also evolved into a modular and adaptable open-source development platform. Its ease of experimentation and extensibility, together with the wide range of available software runtimes and computational kernel libraries, supported its usage as a research tool. This led to the emergence of multiple MemPool \textit{flavors} and, to date, to the tapeout of two chips\footnote{
IIS Chip Gallery (ETH Zürich). \textit{Minpool} (2021), \url{http://asic.ethz.ch/2021/Minpool.html}. \textit{Heartstream} (2024), \url{http://asic.ethz.ch/2024/Heartstream.html}.
}. Such continuous evolution has further expanded the range of trade-offs covered by MemPool while maintaining efficiency and general-purpose applicability within the domain of highly parallel manycore systems.

%%%%%%%%%%%%%%%%%%%%%%%%%%%
% MEMPOOL FLAVORS
%%%%%%%%%%%%%%%%%%%%%%%%%%%

\vspace{-2mm}
\section{MemPool Flavors}
\vspace{-2mm}

Supported by MemPool's open-source, RISC-V-based nature, research into the topic of manycore systems led to the development of many MemPool flavors.
\noindent While MemPool features 256, \textit{TeraPool} scales up to 1024 cores with floating-point support, targeting 6G baseband processing.
\textit{\acrshort{ita} MemPool} (\acrlong{ita} MemPool) focuses on transformer models processing, and \textit{CachePool} explores replacing the shared L1 \gls{spm} with a cache hierarchy. In the following, we compare the main 256-PE variations of MemPool, analyzing their architecture and trade-offs.

\subsubsection{Baseline MemPool}
\begin{figure}[b!]
    \centering
    \includegraphics[width=\columnwidth]{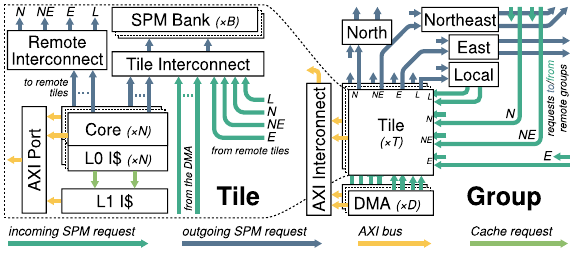}
    \includegraphics[width=\columnwidth]{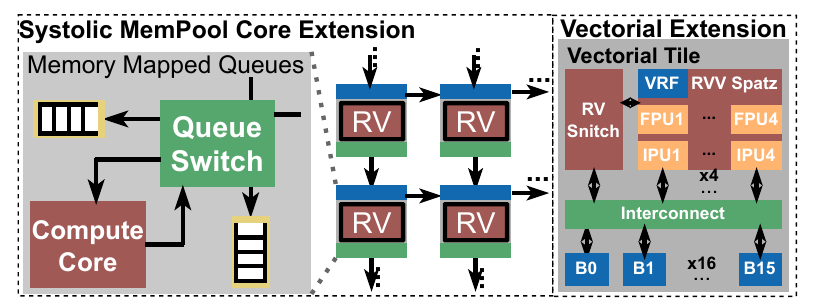}
    \caption{Architecture of the baseline MemPool, also highlighting its systolic and vectorial extensions.}
    \label{fig:tile-arch}
\end{figure}

In its baseline version, MemPool features 256 32-bit Snitch cores with integer arithmetic, \glspl{amo}, and support for custom \gls{dsp} instructions. The 256 cores share 1~MiB of L1 \gls{spm} and have independent instruction paths. \autoref{fig:tile-arch} shows MemPool's hierarchical architecture, where each tile interconnects 4 cores to its 16 \gls{spm} banks with single-cycle latency. 16 tiles form a fully connected group, and 4 groups connect together to compose the MemPool cluster. Each hierarchy level adds two cycles of latency to the memory transactions, easily hidden thanks to the cores' lightweight scoreboard.

% \vspace{-1mm}
\subsubsection{Systolic MemPool}
Many parallel workloads common to telecommunications, artificial intelligence, and image processing feature a highly regular dataflow that can be leveraged to boost performance and energy efficiency.
Systolic MemPool features a low-overhead \gls{isa} extension to implement implicit inter-PE communication and synchronization through L1-mapped queues~[3]. This enables MemPool to configure arbitrary systolic topologies among its \glspl{pe}, without compromising the versatility of the cluster as in systolic-array-based architectures.

% \vspace{-1mm}
\subsubsection{Vectorial MemPool}
Another approach to meeting the intense memory requirement of MemPool's target applications is vector processing. In Vectorial MemPool, a tile only features one scalar 32-bit Snitch core, coupled with a Spatz vector accelerator compliant with the \gls{rvv}~[4]. Each Spatz accelerator includes 4 32-bit \glspl{fpu}.
% Vectorial MemPool results in higher complexity and a reduced number of scalar cores. However,
At the cost of the reduced versatility, Vectorial MemPool leverages the \gls{dlp} of vectorial workloads with \gls{simd} execution, greatly boosting their performance and energy efficiency.

%%%%%%%%%%%%%%%%%%%%%%%%%%%
% RESULTS
%%%%%%%%%%%%%%%%%%%%%%%%%%%

\vspace{-2mm}
\section{Flavor Tasting}
\vspace{-2mm}

\begin{figure}[t]
    \centering
    \includegraphics[width=\columnwidth]{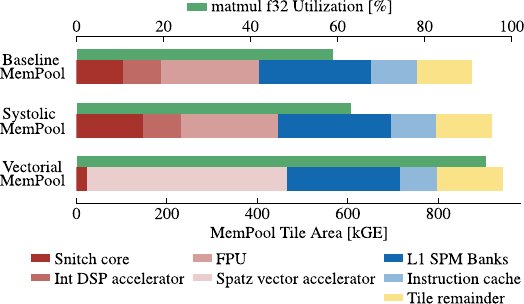}
    \caption{Area breakdown of the MemPool tile in the main MemPool flavors and the corresponding 32-bit floating-point matmul utilization.}
    \label{fig:area-perf}
\end{figure}

\autoref{fig:area-perf} provides an intuition of the wide trade-off range covered by the flavors of MemPool. The area measurements refer to the MemPool's tile implementation in GlobalFoundries’ 12LPP 12~nm advanced FinFET technology. For a proxy of the performance, we employ the utilization of a 32-bit floating-point matrix multiplication kernel.

Baseline MemPool achieves a utilization of 59\%, where the remaining time is spent on control instructions and synchronization.
Systolic MemPool is able to increase the cluster utilization by reducing the amount of explicit load/store instructions. With an area overhead of 5\%, it brings a 7\% improvement in matmul's performance.
With its large vector accelerator, Vectorial MemPool reports the largest tile, with an area 8\% larger than Baseline MemPool. Additional area spent on hardware specialization negatively impacts other generic workloads with higher power consumption. In addition to this, Vectorial MemPool only features one general-purpose core per tile, which decreases its performance with non-vectorizable workloads. However, at the cost of versatility, it achieves up to 94\% utilization with workloads optimized for \gls{rvv}.
All three \textit{flavors} achieve 800MHz@(TT, 0.8V), delivering a peak single-precision floating-point performance of 204.8~GFLOP/s.

%----------------------------------------------------------------------------------------
%	 REFERENCES
%----------------------------------------------------------------------------------------

% \printbibliography % Output the bibliography

\begingroup
\footnotesize % Reduces font size

\endgroup % Ends the smaller font scope

%----------------------------------------------------------------------------------------

\end{document}